\documentclass{aa}  
\usepackage{graphicx,natbib,ulem}
\usepackage{txfonts}
\usepackage{color}
\begin{document}
  \titlerunning{The circumbinary disk of SS 433}
\authorrunning{ Bowler}
   \title{Interpretation of observations of the circumbinary disk of SS 433}

   \subtitle{}

   \author{M. G.\ Bowler \inst{}}

   \offprints{M. G. Bowler \\   \email{m.bowler1@physics.ox.ac.uk}}
   \institute{University of Oxford, Department of Physics, Keble Road,
              Oxford, OX1 3RH, UK}
   \date{Received 1 April 2010 / Accepted 19 July 2010}

 
  \abstract
   {The Galactic microquasar SS\,433 possesses a circumbinary disk most clearly seen in the brilliant Balmer H$\alpha$ emission line. The orbital speed of the glowing material is an important determinant of the mass of the binary system. The circumbinary disk may be fed through the L2 point and in turn may feed a very extended radio feature known as the ruff.} 
  {We present (i) an analysis of spectroscopic optical data from H$\alpha$ and He I spectral lines which reveal the circumbinary disk (ii) comparisons of the rather different signals, to better understand the disk and improve estimates of the rotational speed of the inner rim (iii) a simple model that naturally explains some apparently bizarre spectral variations with orbital phase.}
   {Published spectra, taken almost
     nightly over two orbital periods of the binary system, show H$\alpha$  and He I lines. These were analysed as superpositions of Gaussian components and a simple model in terms of a circumbinary disk was constructed. The possible contributions to the signal of an outflow through the L2 point were considered. }  
  { The data can be understood in terms of a hot spot, generated in proximity to the compact object and rotating round the inner circumbinary disk with a period of 13 days. The glowing material fades with time, quite slowly for the H$\alpha$ source but more rapidly for the He I spectral lines. The orbital speed of the inner rim is approximately 250 km s$^{-1}$. It may be that absorption lines attributed to the atmosphere of the companion are in fact formed in this circumbinary material.}
{ The mass of the binary system must exceed 40 $M_\odot$ and the compact object must be a rather massive stellar black hole. The corollary is that the orbital speed of the companion must exceed 130 km s$^{-1}$.}


   \maketitle
%

\section{Introduction}

The Galactic microquasar  SS 433 is famous for its continual ejection
of plasma in two opposite jets at approximately one quarter the speed
of light. Precession of the jet axis gives rise to the famous moving spectral lines but the so-called stationary lines are more intense. The system is a binary with a period of 13.08 days
(Crampton, Cowley and Hutchings 1980) and eclipses at both oppositions
(e.g. Goranskii et al 1998).  He II 4686 \AA\  emission has been
observed (Crampton and Hutchings 1981,
Fabrika and Bychkova 1990) , attributed to the base of the jets (Fabrika 1997). C II lines orbiting with the compact object have been detected (Gies et al 2002, K. M. Blundell private communication). The orbital speed of the compact object about the binary centre of mass is now well established as 176 km s$^{-1}$ and the mass function as 7.7 $M_\odot$. There is no consistency among the many reports of Doppler speeds for the companion. Relatively recent observations of absorption lines, attributed to the atmosphere of the companion, have yielded an orbital velocity for the companion about the binary centre of mass of 132 km s$^{-1}$ (Cherepashchuk et al 2005), which implies a system mass of 42 $M_\odot$, and 58 km s$^{-1}$ (Hillwig \& Gies 2008, Kubota et al 2010), the latter value implying a system mass of 17 $M_\odot$. Observations  in H$\alpha$ interpreted in terms of a circumbinary disk imply a system mass of approximately 40 $M_\odot$ or greater (Blundell, Bowler \& Schmidtobreick 2008). 

  A rotating ring of glowing gas, viewed almost edge on, will produce a spectrum dominated by radiation from regions to which the line of sight is tangential. A suitable spectral line thus appears split; two horns Doppler shifted by the rotational speed of the ring. Just such a split appears in various stationary features of the spectrum of the microquasar SS 433; Filippenko et al (1988) reported the H Paschen series to be split by approximately 290 km s$^{-1}$ and in the blue various unblended Fe II lines split by 250-300 km s$^{-1}$. Balmer H$\beta$ was split by approximately 480 km s$^{-1}$. These observations covered only a few consecutive days but were interpreted as evidence for a disk structure. The authors evidently thought radiation from the accretion disk most likely, but also considered the possibility of radiation from a circumbinary disk, fed from the L2 point in the SS 433 system. This was taken up by Fabrika (1993) in a prescient paper.
  
    The same two horned structure was observed in H$\alpha$, He I, O I 8446 \AA\ and the Paschen sequence through a campaign of nightly observations of SS 433 with the 3.6-m telescope on La Silla, Chile (Schmidtobreick \& Blundell 2006a,b). The relevant observations covered more than two orbits during a period when SS 433 was quiescent. The two horned signature can be followed night by night in the spectra displayed in Fig.2 of Schmidtobreick \& Blundell (2006b), in both H$\alpha$ and He I. The least noisy signal is to be found in the brilliant H$\alpha$ line and in Blundell, Bowler \& Schmidtobreick (2008) this stationary H$\alpha$ line was fitted as a superposition of Gaussian profiles. It consists of a broad component shown to be associated with the wind from the accretion disk and two narrow components separated by about 400 km s$^{-1}$. These narrow lines run almost railroad straight and do not shift much in position or separation over 30 days (see Fig.1 of Blundell, Bowler \& Schmidtobreick 2008 and Fig.1 of the present paper). The orbital plane of the binary system is almost edge on and this sequence of pairs of narrow lines is the classic signature of the inner rim of a circumbinary disk, radiating strongly all the time from the regions to which the line of sight is tangent. Nonetheless, the intensities vary in antiphase with a period of 13 days, the orbital period of the binary. The interpretation was that the glowing inner rim fades and is refreshed as the binary rotates, possibly through ejection of material through the L2 point or perhaps by ultra violet and X rays from the accretion disk. The observations of SS 433 over an extended period (Blundell, Bowler $\&$ Schmidtobreick 2007,2008) did not reach into the blue, but published data contain in addition to the brilliant Balmer H$\alpha$ line the stronger He I lines at 6678 and 7065 \AA\ , which are also split, have narrow components which fluctuate with opposite phase but fade much faster than H$\alpha$.  Both position and separation contain marked 13 day periodicities, so that it is not immediately obvious that they share the same origin as the simple H$\alpha$ structure. In this paper I present those relevant data and discuss their interpretation in terms of a simple model for the stimulated inner circumbinary disk.  In this model the data are explained by the ring of fire orbiting with a speed of approximately 250 km s$^{-1}$, greater than half the separation of the centres of the fitted Gaussians, the brightest spot rotating close to the compact object and its accretion disk.
    
    There are more recent observations concerning the putative circumbinary disk. First, it has been observed in Brackett $\gamma$ (Perez \& Blundell 2009) over about one orbital period. The extracted rotational velocity is again $\sim$ 200 km s$^{-1}$ but the signal is squeezed between probable accretion disk lines, which complicates its extraction. Secondly, observations in both H$\alpha$ and H$\beta$ suggest that the apparent circumbinary disk lines are not attenuated by the wind from the accretion disk and hence their source is indeed {\it{circumbinary}} (Perez \& Blundell 2010). 
    
    I consider such other possible models for the origin of these split lines as have occurred to me. They are not plausible because of the marked degree to which the red and blue narrow components of H$\alpha$ in Blundell, Bowler \& Schmidtobreick (2008) are unmoving over more than two orbits, which is naturally explained by the disk model.

\section{Data employed in this study}

The relevant spectra were taken nightly from Julian Date 2453000 + 245.5 to +
274.5 and only one observation was missed during this period (Blundell, Bowler \& Schmidtobreick 2008). After JD +274 there are only data at +281 and +282 before another fairly unbroken sequence commenced on JD +287. This was at the onset of an optical outburst, preceding a radio flare, and the stationary lines broadened; an effect attributed to the unveiling of the accretion disk (Bowler 2010). Up to JD +274 H$\alpha$ and He I were usually fitted with three Gaussians, a broad Gaussian (representing an origin in the wind for H$\alpha$) and two narrower.  Where redshifts or Doppler speeds are quoted in this paper, they refer to the centroids of the fitted Gaussians; the relationship between these fitted parameters and the real structure of the source may not be straightforward. The H$\alpha$ data have already been analysed in Blundell, Bowler $\&$ Schmidtobreick (2008) and I have used the results of those published analyses. The evolution with time of the split spectral profiles of H$\alpha$ and of He I at both 6678 and 7065 \AA\ is elegantly presented in Fig.2 of Schmidtobreick  $\&$ Blundell (2006b). In that figure it is immediately obvious that the red and blue components of the split lines alternate in intensity and that the relative intensity in the He I lines varies much more than H$\alpha$. The red side tends to be stronger overall. The results of fitting Gaussian profiles to the He I lines have not been published, so for the purposes of this paper I have made my own fits to the spectra for the He I 6678 \AA\  line, displayed in Fig.2 of Schmidtobreick \& Blundell (2006b). Since this paper involves a comparison of H$\alpha$ and He I, I note here some remarks relevant to the reliability of the fitted parameters in the two cases. As far as H$\alpha$ is concerned, inspection of the top panel of Fig.1 of Blundell, Bowler \& Schmidtobreick (2008) shows that the structure of the line is dominated by a pair of relatively narrow Gaussians sitting on top of a broader component. In Fig.2 of Schmidtobreick \& Blundell (2006b) this can be followed until JD +270, after which a minor component becomes visible in the blue on a few occasions. These additional components are also plotted in the lower panel of that figure. Fitting of two narrow and one broad Gaussian in most cases well represented the spectra; additional terms would either have picked up very minor aspects or have over parametrised the data. Least squares fitting to a complicated shape in terms of many parameters always suffers from the problems of correlated parameters and the existence of local minima in which a fitting program can get trapped; this is not so serious a problem when fitting to three Gaussians as fitting to five, as is necessary after JD +287 (Bowler 2010). Such problems can be dealt with by exploring the parameter space and making independent fits.

  The data for Blundell, Bowler \& Schmidtobreick (2008) were fitted independently with two different least squares programs and for H$\alpha$ the narrow lines seldom differed by more than a Doppler shift of 10 km s$^{-1}$. This is reflected in the random scatter of the results in the lower panel of Fig.1 of Blundell, Bowler \& Schmidtobreick (2008) and the reproduction of those data in Fig.1 of the present paper. The He I spectra are noisier than H$\alpha$ and uncertainties correspondingly larger.
  
  \section{Gaussian fits to the He I 6678 \AA\ line}
  
     Fig.2 of Schmidtobreick $\&$ Blundell (2006b) makes it clear that after JD +287 the H$\alpha$ and He I profiles broaden considerably and become much more complicated. It is not easy to identify Gaussian components  from the circumbinary disk and Blundell, Bowler $\&$ Schmidtobreick (2008) considered only spectra up to JD +274. Similarly, I discuss here only the analysis of  He I  data up to that date. I present results for He I 6678 \AA\ ;  not significantly different from the He I 7065 \AA\ line, as may be seen from Fig.2 of Schmidtobreick \& Blundell (2006b). Fits were made to three Gaussian components in every case, because three were clearly necessary for H$\alpha$, and yielded two narrow components (standard deviation approximately 2 \AA\ ) and a third with standard deviation approximately 7 \AA\ . This third component may not be associated with the wind from the accretion disk because the width does not reflect precession and nodding in the way of the much broader wind component in H$\alpha$ (Blundell, Bowler \& Schmidtobreick 2008). It is possible that the tails represent a lower intensity higher speed source within the rim of the circumbinary disk. The third component does improve the fits but any further additions would certainly overparametrise the data. Here I am concerned only with the signal from the narrow components. The results of fits made in the preparation of Blundell, Bowler \& Schmidtobreick (2008) have not been published, so I digitised the spectra published in Schmidtobreick \& Blundell (2006b), much as I digitised the later period of H$\alpha$ data in Bowler (2010) and made my own fits, the results of which I compared with the earlier work. The random errors on He I narrow components are about 30 km s$^{-1}$, except where a component is both dim and very much out of place, when that component has a Doppler speed uncertain to perhaps 60 km s$^{-1}$. The Doppler speeds of the two narrow Gaussian components fitted to He I 6678 \AA\ are presented in the lower panel of Fig.1 and compared with the data on H$\alpha$ (upper panel) taken from Blundell, Bowler \& Schmidtobreick (2008).The differences readily perceived are certainly real and particularly associated with those days on which red or, about six or seven days later, blue is completely dominant. The differences in the shapes of the raw spectra on such days is obvious in Fig.2 of Schmidtobreick \& Blundell (2006b).
\begin{figure}[htbp]
\begin{center}
{ \includegraphics[width=15cm,trim=0 0 0 2]{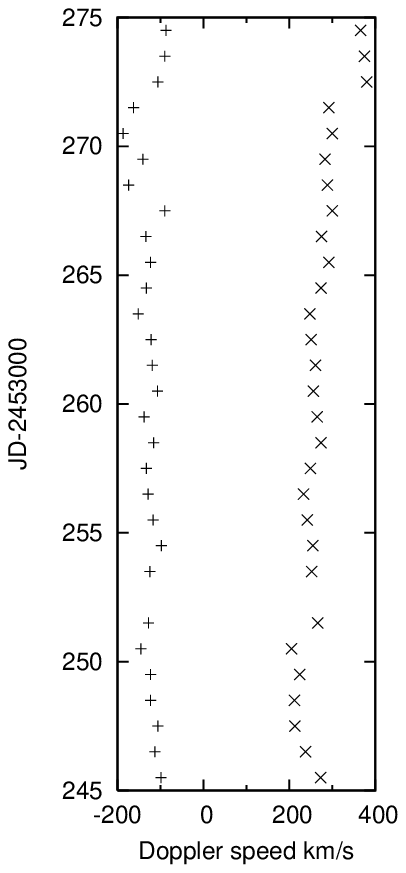} }   
 { \includegraphics[width=15cm,trim=0 0 0 2]{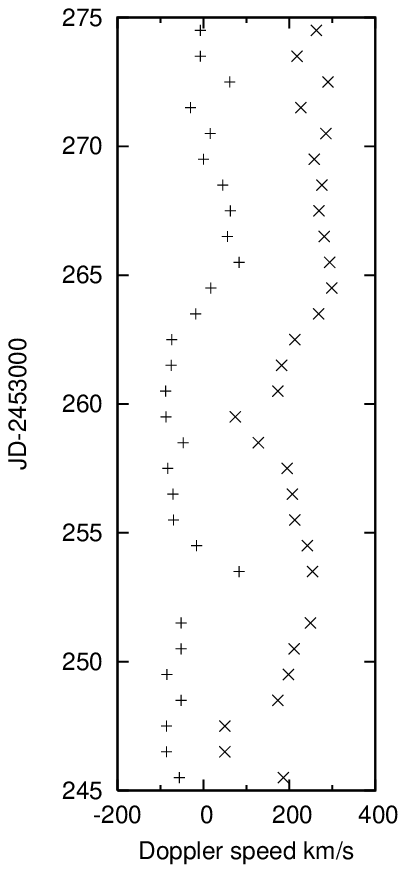} }      
   \caption{Doppler speeds of the blue and red fitted Gaussian components attributed to the circumbinary disk.  Julian date increases vertically. The upper panel is for H$\alpha$ (data from Blundell, Bowler \& Schmidtobreick 2008) and the lower from my fits to He I 6678 \AA\ . In both panels the bluer component is denoted by + and the redder by  x. The He I pattern is bizarre, as discussed in the text. Note in particular the indentations in the blue component at JD +253.5 and in the red at JD +247, 259.5.
   }
\label{fig:ideagram}
\end{center}
\end{figure}

  At first sight, the He I pattern is bizarre; it does not look like the two (almost) straight lines exhibited in the decomposition of H$\alpha$, which provided the classic signature for the existence of the circumbinary disk. Nor is the pattern one of constant spacing oscillating from red to blue, as expected for a circumstellar disk. The blue component of He I closely approaches the red in the vicinities of JD +253 and 266; the red closely approaches the blue in the vicinities of JD + 246 and 259. Nonetheless, this figure is consistent with an origin for the He I lines within the circumbinary disk, for exactly this pattern is generated if radiation from material in the disk fades on a timescale of the order of a few days and is refreshed each orbit by a hot spot rotating with the period of the binary. A very simple model gives remarkable agreement with these data and this is illustrated by a comparison of Fig.1 with the model calculations shown in Fig.2.

\begin{figure}[htbp]
\begin{center}
   \includegraphics[width=15cm,trim=0 0 0 2]{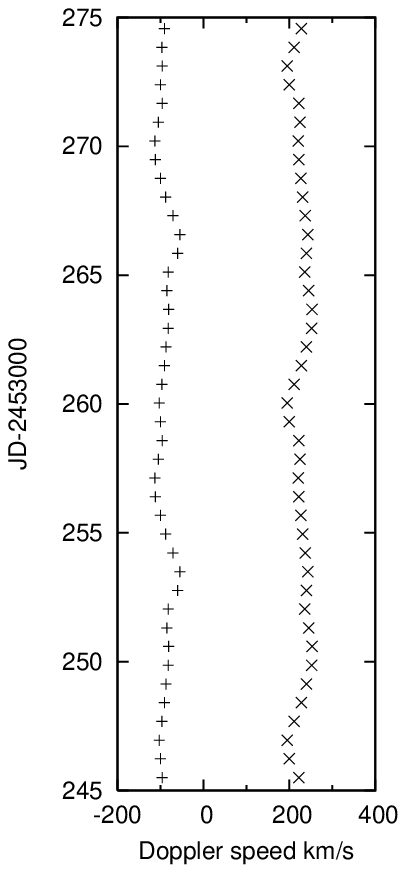} 
   \includegraphics[width=15cm,trim=0 0 0 2]{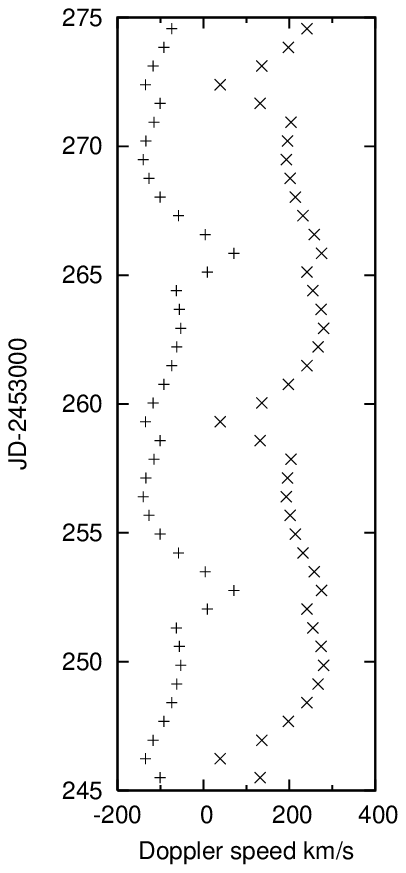}
\caption{ A model calculation for the variation with time of the Doppler speeds of the red and blue Gaussian components. As in Fig.1 + denotes the blue and x the red component. It has been assumed that the circumbinary disk is orbiting the binary system with a speed of 250 km s$^{-1}$, which is (apparently) receding from us at 70 km s$^{-1}$. Radiation from any spot decays exponentially after that spot has been refreshed, with a mean lifetime of one third of an orbital period (lower panel) and one orbital period (upper panel).
 }
\label{fig:disc}
\end{center}
\end{figure}

\section{The model with a circumbinary disk}

The simplest model for radiation from a circumbinary disk is a ring of matter orbiting the binary at a single speed {\it$v$}, transparent to its own radiation. For observation in the plane of the disk, a photon emitted at azimuthal angle $\phi$ is redshifted by a factor 1+ {\it$v/c$} $\sin\phi$, where $\phi$ is $90^{\circ}$ at the tangent where the disk material is receding. If the intensity received from material in the interval d$\phi$ at $\phi$ is

 \begin{equation}
 dI = \rho(\phi)d\phi
 \end{equation}
 
 then the intensity as a function of the Doppler shift is given by
 
 \begin{equation}
  \frac {dI}{dx} = \frac {\rho(\phi)}{\cos\phi} = \frac{\rho(\phi)}{\sqrt{1-x^2}}
 \end{equation}
 
 where {\it$x$} is the Doppler shift in units of the maximum ({\it$v/c$}), equal to $\sin\phi$. Thus if $\rho$($\phi)$ existed only at the line of sight tangents, the spectrum would consist of two lines separated by 2{\it$v/c$}. If $\rho$ is independent of $\phi$ then the spectrum shown in the upper panel of Fig.3 results, horns at both {\it$x$} = +1 and -1 but with significant contributions from those parts of the ring making larger angles with the line of sight. If this elementary form is convoluted with a Gaussian function, some part of which is resolution, the rounded horns move inward and their separation is less than 2{\it$v/c$}.
 
 The model which explains so well the apparently bizarre features of He I (as well as details of the H$\alpha$ spectrum) is a form for $\rho$($\phi)$ which has a leading edge at some value $\phi_0$ and decays exponentially behind that leading edge, winding round the ring until the leading edge is reached again. The whole pattern rotates around the ring with the period of the binary. The lower panel of Fig.3 shows a particular case where the leading edge is located at $\phi_0$=20$^{\circ}$ and the mean decay angle governing the exponential is 0.32 of 2$\pi$ radians. The underlying physical picture is of some hotspot in the circumbinary disk material, rotating with the binary and continually refreshing the glowing material in the ring. For the example shown in the lower panel of Fig.3 the decay time for emission is thus about 4 days. (This is appropriate for He I; for H$\alpha$ the decay time is about 14 days.) This model, having as it does a sharp leading edge, is too simple to match reality, but the final step is to convolve the distribution given by (2) with a Gaussian function. The resolution for these observations is about 0.2 in terms of the relative Doppler shift {\it$x$} and the sharp edges in Fig.3 are removed. The effect of a Gaussian convolution is shown in Fig.4, which shows the convolution of the lower panel of Fig.3 with a Gaussian of standard deviation 0.4. Note that the relative Doppler shift is shown between -2 and +2. The standard deviation of 0.4 is rather larger than resolution, but is supposed to contain some spread in the orbital speeds of the inner rim of the circumbinary disk. It has not been finely tuned but was chosen so that the ratio of separation of the two peaks in H$\alpha$ to the FWHM be about right.
 
   The model was used to calculate the form of the observations shown in Fig.1 (and later figures) in the following way. The predicted convolved spectrum was calculated in terms of the relative Doppler shift {\it$x$} for a number of different values of the decay constant and for every 20$^{\circ}$ in $\phi_0$. The convolved spectra were then fitted with two Gaussians (the amounts, positions and standard deviations all being free parameters); very good representations were obtained in all cases. In Fig.4 is shown not only the convolved spectrum but also the two fitted Gaussians and their sum. It is notable that in this case the centres of both fitted Gaussians are on the blue shifted side; this example corresponds to JD +246.5 and +259.5 in Figs.1 and 2.
   
   The results of these calculations are shown in Fig.2 and may be directly compared with the data in Fig.1.  Figure 2 shows the calculated speeds for the red and blue components, after scaling to an assumed orbital speed for the ring of 250 km s$^{-1}$, with an assumed systemic velocity of 70 km s$^{-1}$ and phase $\phi_0$ equal to zero on JD +245.5. The same parameters are used in Figs. 5 and 6.

\begin{figure}[htbp]
\begin{center}
   \includegraphics[width=9cm]{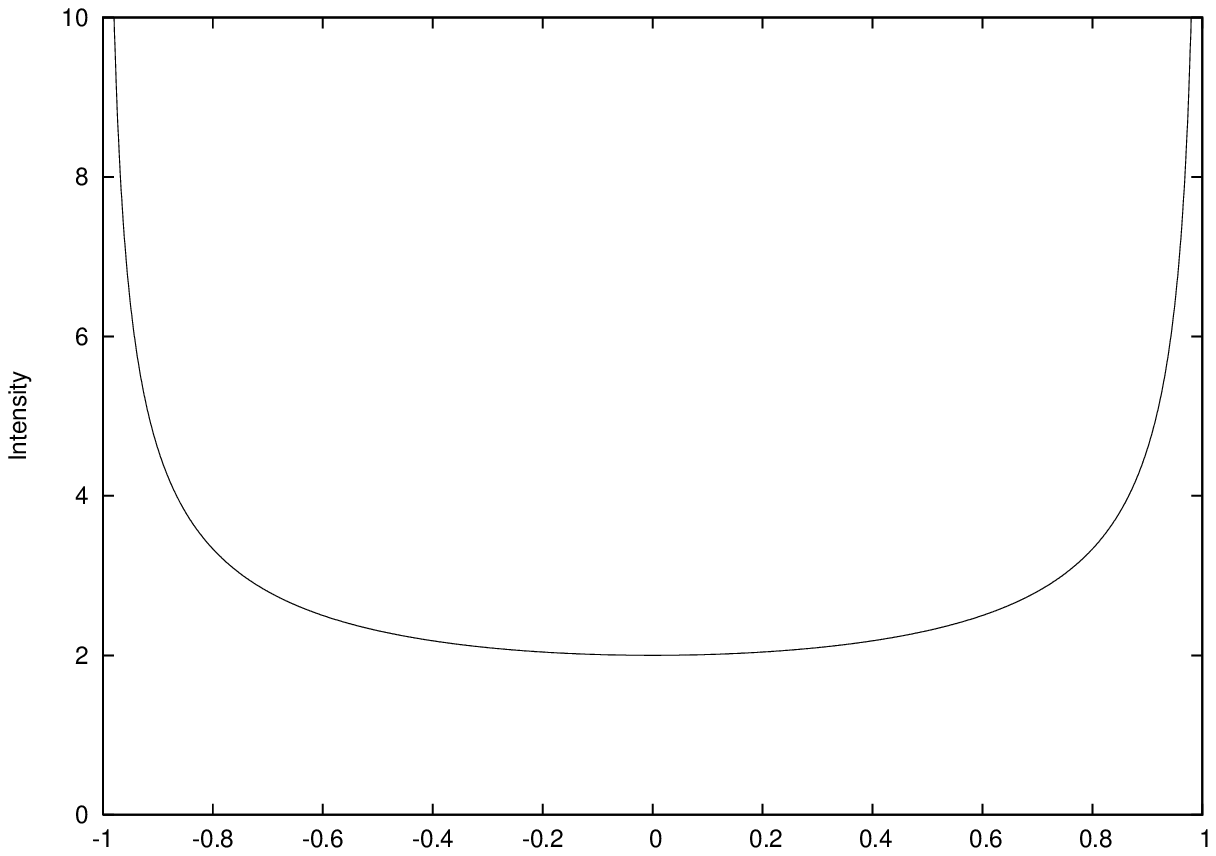} 
   \includegraphics[width=9cm]{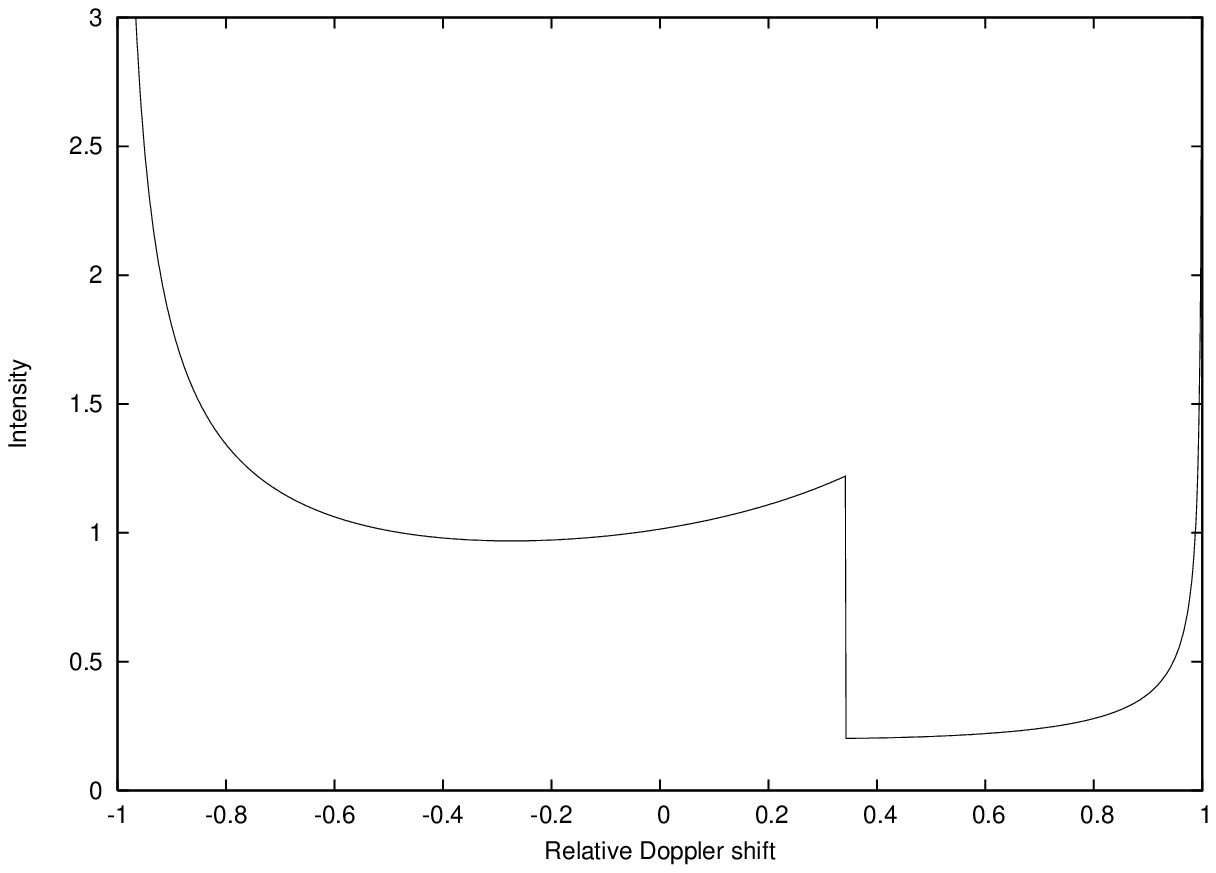}
   \caption{The upper panel shows the intensity of radiation viewed in the plane of a circular ring glowing uniformly all the way round. The ring is transparent and the horns at relative Doppler shifts of -1 and +1 are the result of the concentration of material along the line of sight tangential to the ring. A relative Doppler shift of 1 corresponds to the rotational speed of the ring (Eq.2). The lower panel supposes that a hot spot travels round the ring and stimulated material decays in intensity exponentially, with a mean life of one third of an orbital period. In this example the leading edge is moving into the red, located $20^{\circ}$ after passing relative shift zero. The effect of exponential decay can be seen to the blue. These calculations assume a single rotational velocity and infinite resolution.}
\label{fig:timesequence}
\end{center}
\end{figure}

   For the H$\alpha$ data the railroad track behavior of the two narrow components invited calculation of the rotational speed in terms of  half the difference of the red and blue velocities and the apparent systemic recessional speed from the mean. Both quantities were rather constant with time; Fig. 3 of Blundell, Bowler $\&$ Schmidtobreick (2008), reproduced in Fig.7 of this paper. I have displayed the He I data in the same way, Figs. 5 and 6. Fig.5 shows the nominal rotational speed (half the difference) of the circumbinary disk obtained from He I 6678 \AA\ and from the model as explained above. The agreement between the data and the model is very good. In the model calculation, the maximum (nominal) rotational speed is about 165 km s$^{-1}$, yet the calculation was performed for {\it$v$} equal to 250 km s$^{-1}$. Fig.6 shows apparent systemic recession of the pairs of lines from the fitted Gaussians (half the sum). Again these vary with time; He I agrees very well with the model. Careful examination of Figs. 5 and 6 nonetheless reveals two small discrepancies between the data and the calculation. First, in Fig.5 the maximum (nominal) rotational speed of He in the circumbinary disk is about 145 km s$^{-1}$, corresponding to {\it$v$} equal to 220 km s$^{-1}$ rather than 250 km s$^{-1}$. Secondly, both the rotational and systemic speeds are better reproduced if the phase $\phi_0$ is zero about one day later than JD +245.5.

\begin{figure}[htbp]
\begin{center}
   \includegraphics[width=9cm]{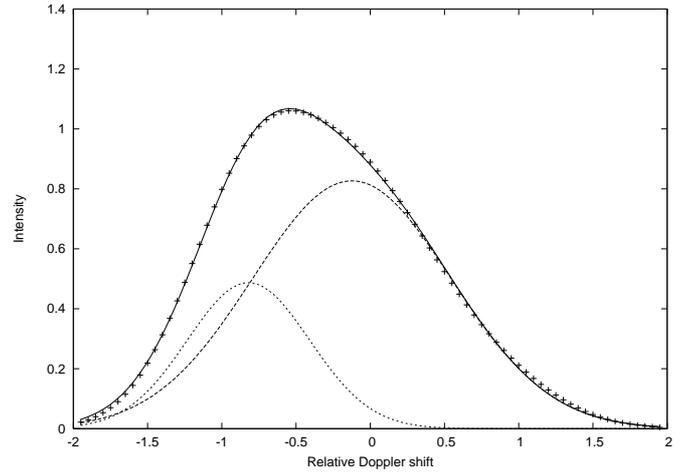} 
   \caption{ The + marks show the spectrum of relative Doppler shift obtained by convolving the distribution shown in the lower panel of Fig.3 with a Gaussian of standard deviation 0.4. This distribution was fitted with two Gaussians, shown individually. The sum is a very accurate representation of the distribution fitted. For this particular phase of the travelling hotspot the centres of both Gaussians are on the negative (blue) side of redshift 0.}
\label{fig:timesequence}
\end{center}
\end{figure}

\begin{figure}[htbp]
\begin{center}
   \includegraphics[width=9cm,trim=0 0 0 140]{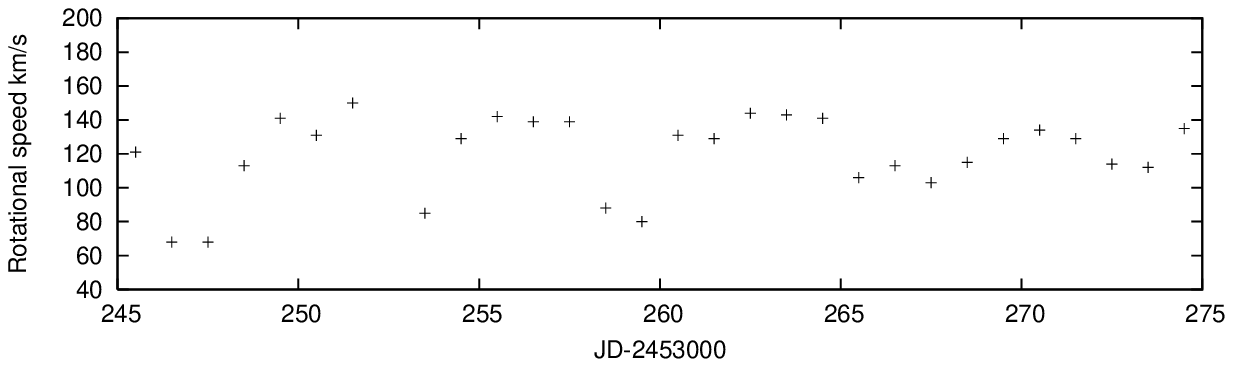}
   \includegraphics[width=9cm,trim=0 0 0 140]{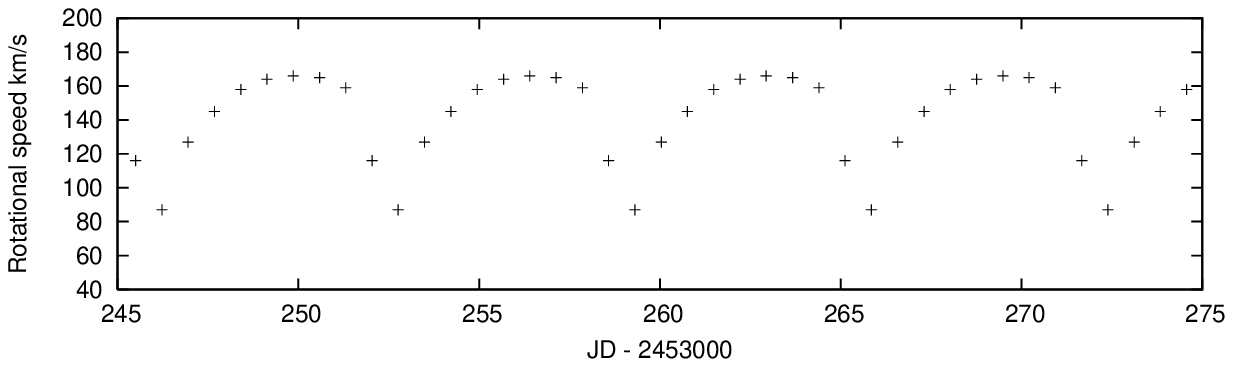}
   \caption{ The nominal rotational velocity of the circumbinary disk, as obtained from the differences between the red and blue components. The top panel is for He I 6678 \AA\ and the bottom panel is the model calculation.}
\label{fig:timesequence}
\end{center}
\end{figure}

\begin{figure}[htbp]
\begin{center}
   \includegraphics[width=9cm,trim=0 0 0 120]{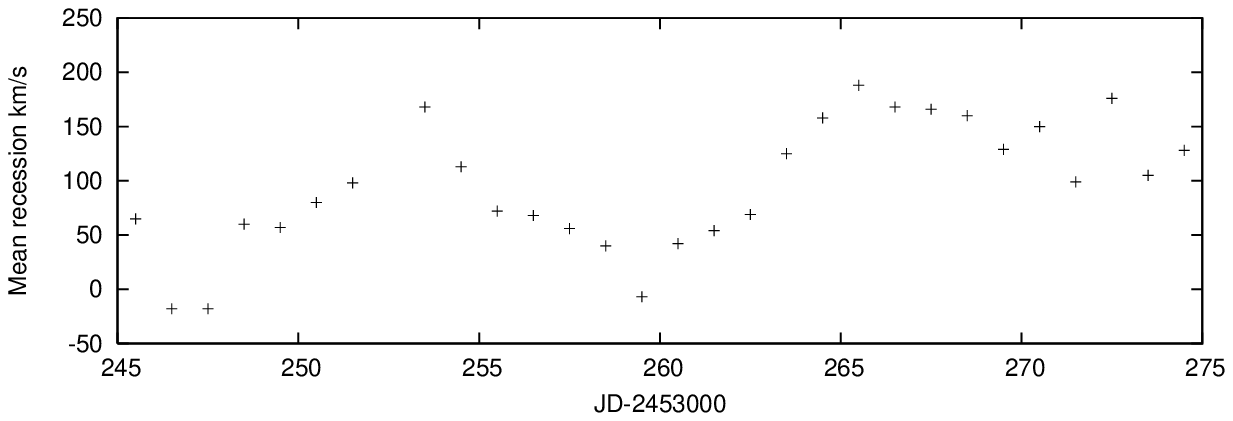}
   \includegraphics[width=9cm,trim=0 0 0 120]{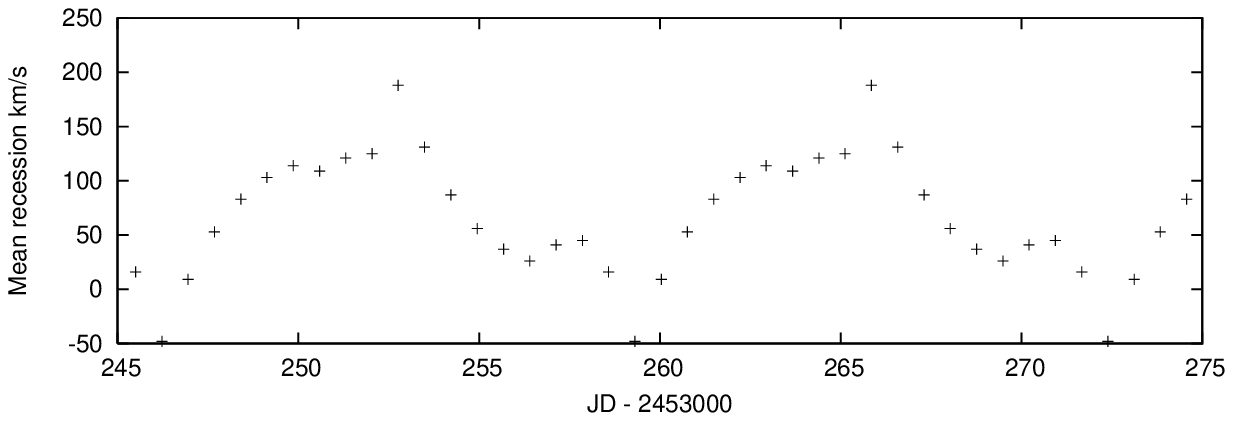}
   \caption{ The mean recessional velocity of the red and blue circumbinary disk components as a function of time. The top panel is for He I 6678 \AA\ and the bottom panel is the model calculation.}
\label{fig:timesequence}
\end{center}
\end{figure}

\section{The model for H$\alpha$}

The original discussion of the H$\alpha$ signature of the circumbinary disk (Blundell, Bowler $\&$ Schmidtobreick 2008) alluded briefly to indentations in the apparent rotational velocity of the circumbinary disk, occurring every half period. Those indentations are much less pronounced than those for He I which are shown in Fig.5 . Examination of the H$\alpha$ and He I lines in Fig.2 of Schmidtobreick $\&$ Blundell (2006b) shows that the He I lines sway from side to side in a much more pronounced fashion than H$\alpha$; in terms of this model that is the result of the decay time of about 4 days. The same model for the fading trail explains the H$\alpha$ data very successfully provided that the H$\alpha$ signal decays with a mean lifetime of about 14 days. Figure 8 has been constructed for direct comparison with Fig.3 of Blundell, Bowler $\&$ Schmidtobreick (2008); for convenience those data  are reproduced here in Fig.7. The decay time is 13.9 days, the rotational speed of the ring 250 km s$^{-1}$ and the assumed systemic velocity 70 km s$^{-1}$. The upper panels of Figs.7 and 8 show the rotational and systemic speeds as calculated from the sum and difference velocities; the lower panel shows the variation with time of the difference between the areas of the red and blue H$\alpha$ components, as calculated from the Gaussians fitted to the data and to the model spectra. The model does not contain the tendency for the red side of the circumbinary disk  to dominate.

 The agreement between Fig.3 of Blundell, Bowler $\&$ Schmidtobreick (2008) and Fig.8 is good, in that very similar structure is present, as would be expected from a comparison of Fig.1 with Fig.2. The model with the slowly fading tail explains the indentations every half period noticed in the rotational speeds and also indentations in the systemic speed. A careful comparison suggests first that the speed of the ring which glows in H$\alpha$ might be better represented by a value greater than 250 km s$^{-1}$ by 20 or 30 km s$^{-1}$ (the H$\alpha$ data show a slow increase in apparent rotational speed over time, not included in the model) and secondly that the data might be better explained if the leading edge is at phase $\phi_0$ = 0 on JD + 248 rather than 245.5. This is about a day later than for He I. The phase of the binary orbit, according to the Goranskii ephemeris [Goranskii, Esipov $\&$ Cherepaschuk  (1998)] , is 0.46 at JD +248, corresponding to onset of eclipse of the companion by the disk of the compact object.
 
 \begin{figure}[htbp]
 \begin{center}
    \includegraphics[width=9cm,trim=0 0 0 100]{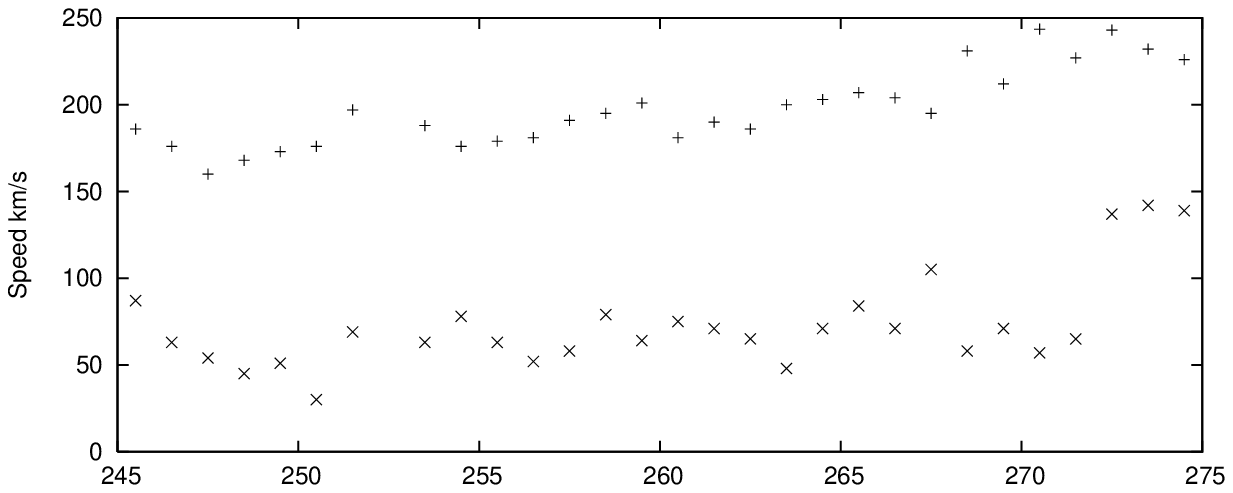}
    \includegraphics[width=9cm,trim=0 0 0 100]{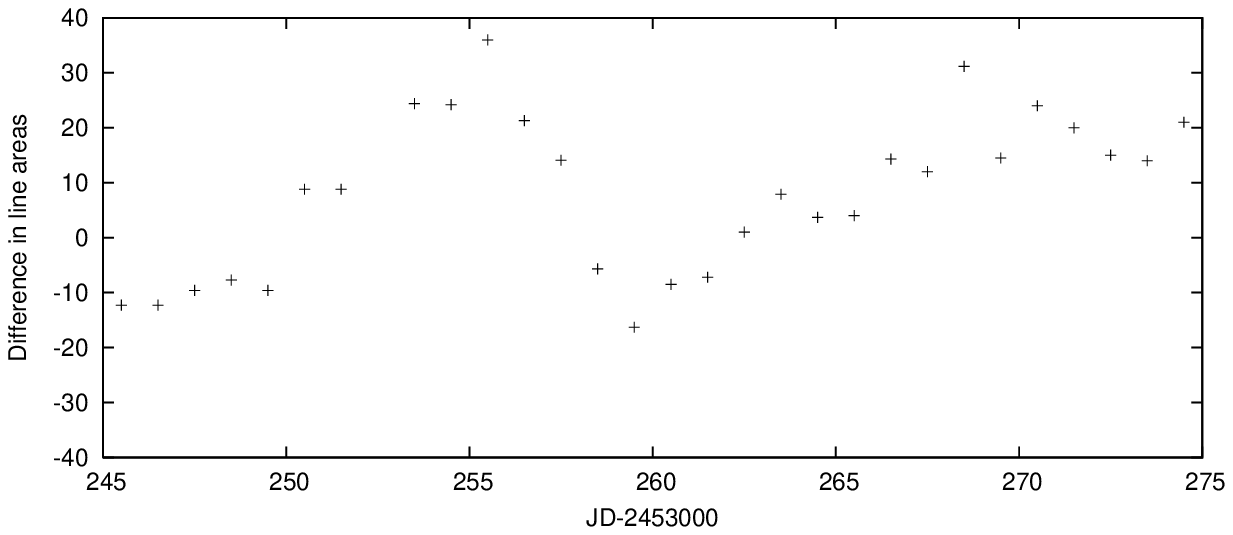}
    \caption{ H$\alpha$ data from Blundell, Bowler \& Schmidtobreick (2008). The upper panel shows rotational speed as + and the apparent systemic recessional velocity as x. The lower panel shows the alternating difference in areas of the red and blue lines.}
 \label{fig:timesequence}
 \end{center}
 \end{figure}

\begin{figure}[htbp]
\begin{center}
   \includegraphics[width=9cm,trim=0 0 0 100]{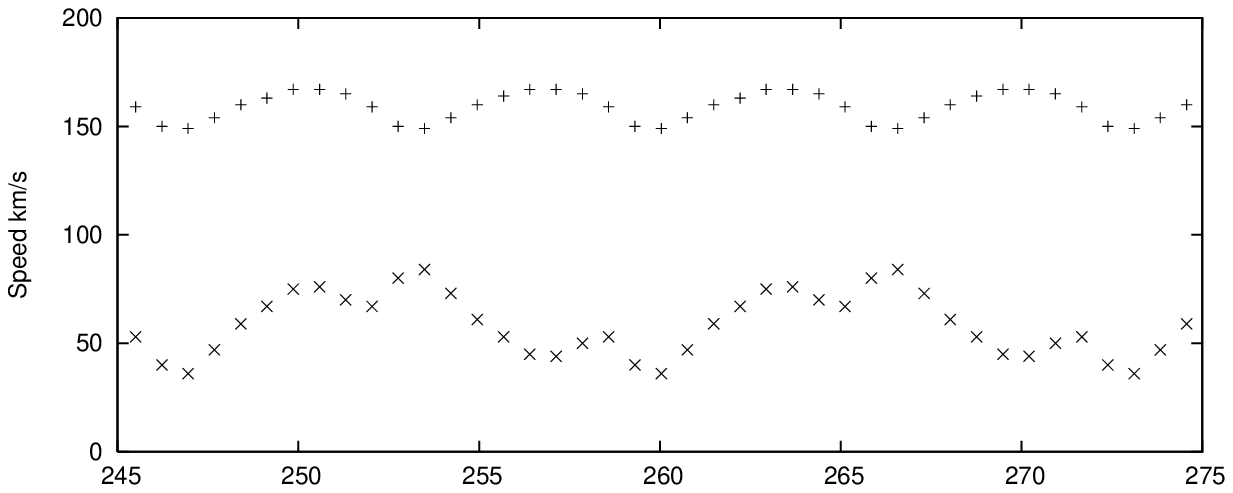} 
   \includegraphics[width=9cm,trim=0 0 0 100]{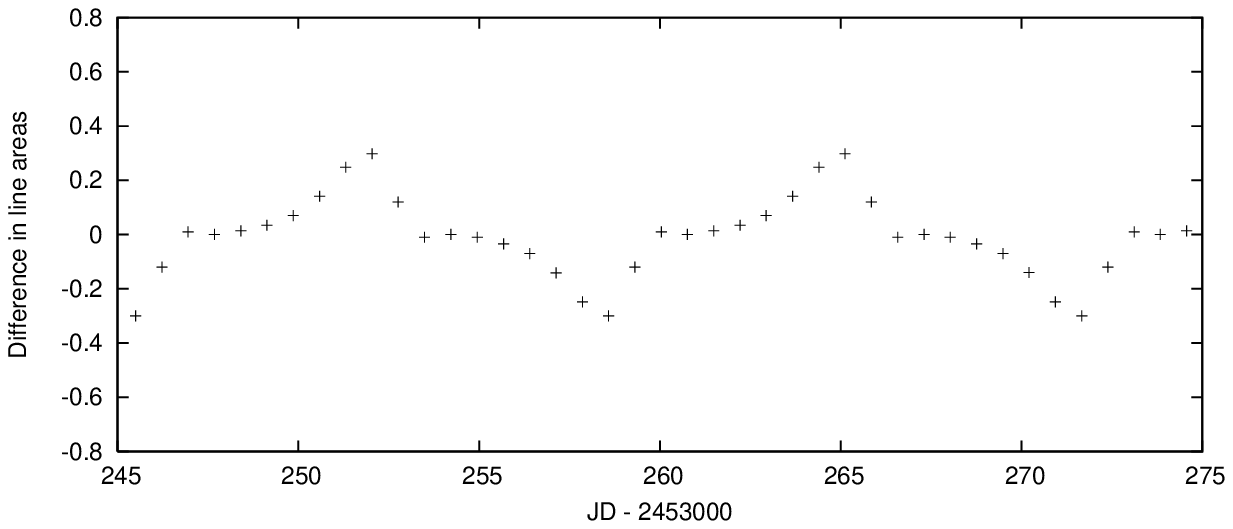}
   \caption{ Model calculations for a slowly fading trail following the hotspot, to be compared with H$\alpha$ data. The upper panel shows rotational speed as + and the apparent systemic recessional velocity as x. The lower panel shows the alternating difference in areas of the red and blue lines.}
\label{fig:timesequence}
\end{center}
\end{figure}

\section{Are there alternative explanations?}

   The distribution shown in Fig.3 (upper panel) and derived from Eq.2 with an azimuthally symmetric intensity is also obtained for a ring expanding uniformly outwards with speed{ \it{v}.} Thus every aspect of the models developed above would be reproduced by a ring expanding radially with constant speed {\it{v}}  of about 250 km s$^{-1}$. This configuration would have to be maintained over weeks at least with constant speed of expansion and pretty much constant intensity. This is not plausible. However, an expanding spiral of material ejected from the L2 point would be initially approximately tangential to the binary orbit but the motion rapidly becomes largely radial. This is the configuration envisaged by Fabrika (1993) and it is not only plausible but is supported by evidence from P Cygni like absorption features observed on the blue side of (among others) H$\alpha$ and He I emission. These absorption lines correspond to an equatorial outflow with a speed of $\sim$ 200 km s$^{-1}$, most visible when the accretion disk is edge on to the line of sight. The speed of outflow along the line of sight oscillates with the orbital period; Kopylov et al (1989) report for H$\alpha$ a mean radial velocity of $ -$290 km s$^{-1}$ with an amplitude of $\sim$ 185 km s$^{-1}$; there is considerable scatter. They further suggest that the 13 day period oscillation is because the ejected gas slows down as it draws further away from the centre of mass of the binary. The narrow H$\alpha$ lines attributed to a circumbinary disk have a 13 day oscillation with an amplitude of no more than 25 km s$^{-1}$, attributed in the model of section 5 to a slow fading of the emission with orbital phase.

       The flow of gas from the companion to a compact object in a close binary has been modelled by Sawada et al (1986). These authors followed overflow gas as far as the L2 point and concluded that a substantial fraction of gas from the companion is discarded through the L2 point (but not through the L3 point). I am not aware of any calculations exploring in detail its subsequent fate. Some insight can however be gained from a rather simple calculation. The radius from the centre of mass of the binary of the L2 point is, as a fraction of the binary separation {\it{A}}, almost independent of the ratio of the mass of the compact object to that of the companion, {\it{q}}. It ranges from a value of $1.26A$ for $q=0.35$ to $1.20A$ for $q=1$. The orbital speed of the L2 point is then given by a fraction $1.2(1+q)$ of the orbital velocity of the compact object and is in all cases of interest a little smaller than the escape velocity from the L2 radius. Thus one may expect some spillover from L2 to leave the binary system eventually and other material to fall back and either be captured or join a stable circumbinary ring. It is simple to follow (at least approximately) the fate of material leaving the L2 point tangentially at just the escape velocity, along a parabolic orbit. It reaches the inner stable circumbinary radius ($\sim 2A$) after 4 days, making an angle of approximately 45$^{o}$ and moving at 1.25 the speed of that orbit. Because of the rotation of the whole system, matter moving in the opposite direction to the material just leaving L2 was launched about 8 days earlier and has reached a radius of $3.4A$. At that distance it is now moving with half the speed of material leaving the L2 point tangentially. Thus at orbital phase 0.75, when the L2 point is receding at maximum line of sight velocity, the approaching material on the far side of the spiral structure is moving, in the binary centre of mass, at half the L2 speed. It has also probably faded very considerably. If the two extreme components could be properly identified either of the two narrow lines would oscillate with a period of 13 days and amplitude about one third of the mean Doppler speed. At least in the absence of a detailed model, it would not be possible to rule out an expanding spiral structure as the origin of the He I lines, but the above considerations do not seem consistent with the extreme stability of H$\alpha$.

\section{Conclusions}

     The great stability of the narrow red and blue components of the stationary H$\alpha$ line in the spectrum of SS 433 is easily understood in terms of an orbiting circumbinary ring, presumed to be the inner rim of a larger disk. This stability is very much at odds with a source in an outward flow through the L2 point. The more complicated behaviour of the He I lines is consistent with an origin in the circumbinary disk, provided only that the more rapid fading can be accomodated. Thus the very simple model in which radiation from the circumbinary disk decays exponentially behind a leading edge, convoluted with a Gaussian function, accounts astonishingly well for the narrow components found within the stationary H$\alpha$ and He I lines. H$\alpha$ at least is contributed by radiation from the inner circumbinary disk, orbiting the binary at very approximately 250 km s$^{-1}$. The apparent systemic velocity of the ring is approximately 70 km s$^{-1}$.  H$\alpha$ emission fades on a timescale of 14 days whereas He I has a fading time of about 4 days. In this simple model the leading edge of the H$\alpha$ emission is found close to the passage of the compact object and its disk but the leading edge for He I is perhaps a day or so earlier. The irradiation of a given point on the circumbinary disk by the source of intense radiation in the vicinity of the compact object varies by a factor of almost three over the orbit, from geometric effects alone. When the compact object is furthest from a point on the disk, additionally the companion eclipses that portion. This suggests that intense radiation from the vicinity of the compact object periodically refreshes emission from the circumbinary disk - the effect might be augmented by arrival of material from L2 - and a decay scale of about one period is not unreasonable. It is perhaps curious that the He I signal decays much faster than H$\alpha$ and the hot spot is ahead of the compact object. It seems entirely possible that some phases at least of the He I lines are dominated by radiation from the stream leaving the L2 point, which feeds the circumbinary disk and eventually the wider environment. For the remainder of this paper I shall suppose that the circumbinary disk is real and that the inner edge has at least approximately the orbital speed extracted from the model. The remaining uncertainty is the radius at which the ring of fire orbits.

The rotational speed of the inner circumbinary disk provides an important constraint on the mass of the system and hence on the mass of the compact object. If the radius at which the material orbits with speed $v$ is $fA$, $A$ being the separation of the two members of the binary, then the mass of the system {\it$M_{\rm S}$} is given by

\begin{equation}
M_S = 1.35 f^{3/2}(v/100)^3
\end{equation}

in units of {\it$M_{\odot}$}, {\it$v$} being specified in km s$^{-1}$ [Eq. 3 of Blundell, Bowler $\&$ Schmidtobreick (2008)]. The innermost stable orbit about the binary system corresponds to {\it$f$} approximately 2, estimates varying between 1.8 and 2.3. However, in a system such as SS 433 where material is almost certainly being added to the circumbinary disk via the L2 point, the glowing inner rim may be within the innermost stable orbit, depending on the residence time and the rate at which matter is added. I show in Table 1 the mass of the system {\it$M_S$}, the companion {\it$M_C$} and the compact object (including the mass of the accretion disk) {\it$m_X$} as a function of the value of {\it$f$}.

\begin{table}
\centering
\vspace{1cm}
\begin{tabular}{llrr}
\hline
 $f$        &$M_{\rm S}$       &$M_{\rm C}$        &$m_{\rm X}$  \\
  1.5       &38.8                     &22.6                       &16.1                \\
  1.6       &42.7                     &24.1                       &18.6                \\
  1.7       &46.8                     &25.6                       &21.0                \\
  1.8       &51.0                     &27.2                       &23.9                \\
  1.9       &55.3                     &28.7                       &26.7                \\
  2.0       &59.7                     &30.2                       &29.6                \\
  2.1       &64.2                     &31.7                       &32.6               \\
  2.2       &68.8                     &33.2                       &35.8               \\
  2.3       &73.6                     &34.7                       &38.9                \\
 
 \end{tabular}
 \caption{\label{tab:two} Masses in the binary system SS 433 as a function of the radius parameter $f$. $M_{\rm S}$ is the total mass, $M_{\rm C}$ is the mass of the companion and $m_{\rm X}$ is the mass of the compact object, in units of $M_\odot$ . ( From Eq.3, assuming $\it v $ equal to 250 km s$^{-1}$. The masses $M_{\rm S}$ scale with the cube of the orbital speed.)}
 \end{table}
 
Table 1 has been composed assuming that the material in the ring of fire is orbiting the centre of mass of the binary at 250 km s$^{-1}$. This value is model dependent; to the extent that the model is accurate the radiating He is orbiting slower than H. It is most unlikely that bulk material radiating in the inner rim of the circumbinary disk is orbiting slower than 200 km s$^{-1}$. For this speed, the system mass is 40 $M_\odot$ for a value of $f$ of 2.3, as assumed in Blundell, Bowler \& Schmidtobreick (2008); material orbiting in the ring of fire at 250 km s$^{-1}$ would lie further inwards, $f$ of 1.5. Perhaps the H$\alpha$ radiation is coming from increasing amounts of material close to joining the circumbinary disk rather than in stable orbits. In any event, the evidence from the circumbinary disk is that the system is massive, almost certainly exceeding 40 $M_{\odot}$, and the compact object is a rather massive stellar black hole.

     The Doppler shifts of a number of lines reported by Cherepashchuk et al. (2005) and attributed to the companion yield an orbital velocity of the companion about the binary centre of mass of 132 km s$^{-1}$ and hence a system mass of 42 $M_{\odot}$. This is not in disagreement with data on the circumbinary disk, although $f$ would have to be a little less than 1.7 for disk material moving as fast as 250 km s$^{-1}$. The observations of Hillwig $\&$ Gies (2008), Kubota et al (2010), interpreted as absorption in the atmosphere of the companion, are not consistent with the circumbinary disk modelled in this paper. They infer an orbital velocity for the companion of 58 km s$^{-1}$ and hence a system mass of about 17 M$_{\odot}$. For disk material orbiting at 250 km s$^{-1}$ this yields a value of $f$ of 0.85 - unbelievably close in because the L2 radius corresponds to $f=1.26$. (For material at 220 km s$^{-1}$ $f$ would be 1.1 and even for 200 km s$^{-1}$ an $f$ of 1.33.) However, it seems possible that the absorption lines of Hillwig \& Gies (2008) and Kubota et al (2010) are in fact produced in circumbinary material. The arguments against a circumbinary origin for such absorption lines set out in Hillwig et al (2004) do not obviously apply to material in an orbiting disk. The observations of a sinusoidal oscillation of amplitude $\sim$ 60 km s$^{-1}$, sharing the orbital phase of the companion, are explained quantitatively if the origin of these lines is absorption of continuum light from the companion in circumbinary material orbiting at $\sim$ 240 km s$^{-1}$. The reason is that the companion presents an orbital radius projected on the sky of $R_{\rm C}{\sin\phi}$, where here $\phi$ is the orbital phase, and at this elongation is viewed through disk material moving with a radial component of velocity $V_{\rm r}$ given by 
   
   \begin{equation}
   V_ r = \frac{R_C}{R_D}V_D{\sin\phi}
    \end{equation}
    
     where the radius at which the circumbinary material orbits at speed $V_D$ is $R_D$. The radius of the companion orbit is $\sim A/2$ and the disk material orbits at $\sim 2A$. This is most easily illustrated by considering the configuration at an orbital phase of 0.25, when the companion is receding. At extreme elongation any circumbinary material through which it  is seen has, in the centre of mass system of the binary, a recessional velocity of $\sim$ 60 km s$^{-1}$. Perhaps it is just a coincidence, but the numerical agreement is as good as it could be. The absorption line data of Hillwig \& Gies (2008) and Kubota et al (2010) might thus be reconciled with the data attributed to the circumbinary disk which, as analysed in this paper, imply an orbital velocity for the companion exceeding 130 km s$^{-1}$. If those absorption lines do in fact originate in the circumbinary disk then identification of the spectral type of the companion as mid A is premature.

\begin{acknowledgements}
I had many stimulating discussions with K. M. Blundell, who carried out extremely painstaking work on analysis of stationary lines in terms of Gaussian components, during the preparation of Blundell, Bowler \& Schmidtobreick (2008). I thank an anonymous referee for a number of fruitful comments.
\end{acknowledgements}

\end{document}